\documentclass[11pt]{article}

\newsavebox{\foobox}
\newcommand{\slantbox}[2][0]{\mbox{%
        \sbox{\foobox}{#2}%
        \hskip\wd\foobox
        \pdfsave
        \pdfsetmatrix{1 0 #1 1}%
        \llap{\usebox{\foobox}}%
        \pdfrestore
}}
\newcommand\unslant[2][-.25]{\slantbox[#1]{$#2$}}

\newcommand{\mpi}{\text{\unslant[-.18]\pi}}
\newcommand{\mdelta}{\text{\unslant[-.18]\delta}}

\usepackage[left=2cm, right=2cm, top=2.5cm, bottom=2.5cm]{geometry}
\usepackage[x11names]{xcolor}
\usepackage{fancyhdr, amssymb, cancel, amsmath, graphicx, pgfplots, tikz, dsfont}
\usepackage{isomath,bbold}

\usetikzlibrary{shadows}

\newcommand{\stylecolor}{blue!50!black}

\usepackage[labelfont={bf,sf, color=\stylecolor}, margin={1.5cm,0cm}]{caption}

\usepackage[colorlinks=true, urlcolor=\stylecolor!70!white, linkcolor=\stylecolor, citecolor=\stylecolor!70!white, hyperindex=true, linktocpage=true]{hyperref}

\usepackage[explicit]{titlesec}

\newcommand*\sectionlabel{}
\titleformat{\section}
  {\gdef\sectionlabel{}
   \Large\bfseries\scshape}
  {\gdef\sectionlabel{\thesection }}{0pt}
  {\begin{tikzpicture}[remember picture,overlay]
       \end{tikzpicture}
  }
\titlespacing*{\section}{0pt}{0pt}{0pt}

\newcommand*\subsectionlabel{}
\titleformat{\subsection}
  {\gdef\subsectionlabel{}
   \large\bfseries\scshape}
  {\gdef\subsectionlabel{\thesubsection  }}{0pt}
  {\begin{tikzpicture}[remember picture]
    	\draw (-0.15, 0) node[left] {\color{\stylecolor} \textsf{\subsectionlabel}};
	\draw (0.15, 0) node[right] {\color{\stylecolor} \textsf{#1}};
	\fill[color=\stylecolor] (-0.05, -0.23) rectangle (0.05, 0.23);
       \end{tikzpicture}
  }
\titlespacing*{\subsection}{-4pt}{10pt}{0pt}

\newcommand*\subsubsectionlabel{}
\titleformat{\subsubsection}
  {\gdef\subsubsectionlabel{}
   \bfseries\scshape}
  {\gdef\subsubsectionlabel{\thesubsubsection.\ \  }}{0pt}
  {\begin{tikzpicture}[remember picture]
    	\draw (0, 0) node[left] {\color{\stylecolor} \textsf{\subsubsectionlabel}};
	\draw (0, 0) node[right] {\color{\stylecolor} \textsf{#1}};
       \end{tikzpicture}
  }
\titlespacing*{\subsubsection}{-4pt}{7pt}{0pt}

\pgfplotsset{every axis legend/.append style={at={(1.02,1)},anchor=north west}}

\begin{document}

\allowdisplaybreaks

\pagestyle{fancy}
\renewcommand{\headrulewidth}{0pt}
\fancyhead{}

\fancyfoot{}
\fancyfoot[C] {\textsf{\textbf{\thepage}}}

\begin{equation*}
\begin{tikzpicture}
\draw (\textwidth, 0) node[text width = \textwidth, right] {\color{white} easter egg};
\end{tikzpicture}
\end{equation*}

\begin{equation*}
\begin{tikzpicture}
\draw (0.5\textwidth, -3) node[text width = \textwidth] {\huge  \textsf{\textbf{Sign of viscous magnetoresistance in electron fluids}} };
\end{tikzpicture}
\end{equation*}
\begin{equation*}
\begin{tikzpicture}
\draw (0.5\textwidth, 0.1) node[text width=\textwidth] {\large \color{black}  \textsf{Ipsita Mandal}$^{\color{\stylecolor} \textsf{a,b}}$ \textsf{ and Andrew Lucas}$^{\color{\stylecolor} \textsf{c,d}}$ };
\draw (0.5\textwidth, -0.5) node[text width=\textwidth] {\small $^{\color{\stylecolor} \textsf{a}}$ \textsf{Laboratory of Atomic And Solid State Physics, Cornell University, Ithaca, NY 14853, USA}};
\draw (0.5\textwidth, -1) node[text width=\textwidth] {\small $^{\color{\stylecolor} \textsf{b}}$ \textsf{Faculty of Science and Technology, University of Stavanger, 4036 Stavanger, Norway}};
\draw (0.5\textwidth, -1.5) node[text width=\textwidth] {\small $^{\color{\stylecolor} \textsf{c}}$ \textsf{Department of Physics, Stanford University, Stanford, CA 94305, USA}};
\draw (0.5\textwidth, -2) node[text width=\textwidth] {\small $^{\color{\stylecolor} \textsf{d}}$ \textsf{Department of Physics, University of Colorado, Boulder, CO 80309, USA}};
\end{tikzpicture}
\end{equation*}
\begin{equation*}
\begin{tikzpicture}
\draw (0, -13.1) node[right, text width=0.5\paperwidth] 
{\texttt{ipsita.mandal@gmail.com, andrew.j.lucas@colorado.edu}};
\draw (\textwidth, -13.1) node[left] {\textsf{January 22, 2020}};
\end{tikzpicture}
\end{equation*}
\begin{equation*}
\begin{tikzpicture}
\draw[very thick, color=\stylecolor] (0.0\textwidth, -5.75) -- (0.99\textwidth, -5.75);
\draw (0.12\textwidth, -6.25) node[left] {\color{\stylecolor}  \textsf{\textbf{Abstract:}}};
\draw (0.53\textwidth, -6) node[below, text width=0.8\textwidth, text justified] {\small In sufficiently clean metals, it is possible for electrons to collectively flow as a viscous fluid at finite temperature.  These viscous effects have been predicted to give a notable magnetoresistance, but whether the magnetoresistance is positive or negative has been debated.  We argue that regardless of the strength of inhomogeneity, bulk magnetoresistance is always positive in the hydrodynamic regime.  We also compute transport in weakly inhomogeneous metals across the ballistic-to-hydrodynamic crossover, where we also find positive magnetoresistance.  The non-monotonic temperature dependence of resistivity in this regime (a bulk Gurzhi effect) rapidly disappears upon turning on any finite magnetic field, suggesting that magnetotransport is a simple test for viscous effects in bulk transport, including at the onset of the hydrodynamic regime.};
\end{tikzpicture}
\end{equation*}

\tableofcontents

\begin{equation*}
\begin{tikzpicture}
\draw[very thick, color=\stylecolor] (0.0\textwidth, -5.75) -- (0.99\textwidth, -5.75);
\end{tikzpicture}
\end{equation*}

\titleformat{\section}
  {\gdef\sectionlabel{}
   \Large\bfseries\scshape}
  {\gdef\sectionlabel{\thesection }}{0pt}
  {\begin{tikzpicture}[remember picture]
	\draw (0.2, 0) node[right] {\color{\stylecolor} \textsf{#1}};
	\draw (0.0, 0) node[left, fill=\stylecolor,minimum height=0.27in, minimum width=0.27in] {\color{white} \textsf{\sectionlabel}};
       \end{tikzpicture}
  }
\titlespacing*{\section}{0pt}{20pt}{5pt}

\section{Introduction}
In an ultrapure solid-state device, it may be the case that momentum-conserving electron-electron scattering is the fastest process which can scatter an electronic quasiparticle \cite{gurzhi}.  Historically such metals did not exist: in a Fermi liquid, the electron-impurity scattering rate is always faster as temperature $T\rightarrow 0$, and at higher $T$ usually an umklapp process (off electrons or phonons) is sufficient to relax the electronic momentum.   Nevertheless, experiments have increasingly discovered evidence for this hydrodynamic flow regime in clean samples of graphene, GaAs, and other compounds, in recent years \cite{molenkamp, bandurin, crossno, levitov1703, bakarov, bandurin18, walsworth, sulpizio}; see \cite{lucasreview} for a review.

While there are possible applications for hydrodynamic electron flow ranging from high conductance mesoscopic devices \cite{levitov1607} to terahertz radiation generation \cite{DS}, this paper is inspired by a simpler question: is it possible that hydrodynamic effects have already been seen in experiments via the unusual behavior of electrical resistivity as a function of temperature, etc.?   A number of clear predictions have already been made for the resistivity of a Fermi liquid of viscous electrons \cite{andreev, lucas3, hartnoll1706, lucas1711}, but there is no compelling experimental observation thus far.

It was recently suggested \cite{alekseev} that viscous electron flow through an inhomogeneous device would lead to negative magnetoresistance in a (quasi-)two-dimensional metal, where the dissipative resistivity decreases as one turns on a small magnetic field:  $\partial \rho / \partial (B^2) < 0$.  {\color{black} This work was inspired by an experiment on relatively large samples of GaAs \cite{kwwest}, where negative magnetoresistance has been observed, albeit not necessarily in a hydrodynamic regime.  More recently, negative magnetoresistance was seen in very narrow channels of graphene at moderately high temperatures (where hydrodynamic effects are observed) \cite{bandurin19}.}   Later, \cite{levchenko1, levchenko2} pointed out that in bulk crystals, the magnetoresistance would always be positive; however, their argument is perturbative in the strength of the inhomogeneity. 

In this paper, we will argue that the conclusion of \cite{levchenko1, levchenko2} is valid more generally, and thus that negative magnetoresistance is \emph{not} a signature of viscous flow in a bulk crystal.  We do so from two perspectives.  First, we will argue that even when disorder and inhomogeneity are arbitrarily strong, a generic Fermi liquid will exhibit positive magnetoresistance.  Our conclusion is based on an exact analysis of the hydrodynamic transport problem in systems which are homogeneous in one out of the two spatial dimensions, along with a discussion of the density dependence of the local hydrodynamic coefficients.    Secondly, we will argue that near a low temperature transition between viscous electron flow and ballistic electron flow, there is an enormous positive magnetoresistance in weakly inhomogeneous systems.  Therefore, the conclusion of \cite{levchenko1, levchenko2} cannot be avoided by studying systems near the onset of viscous flow.  In fact, we argue that the magnetic field dependence of resistivity is so strong that magnetoresistance is an excellent test for whether resistance minima (where $\rho(T)$ is a decreasing function at low $T$) are a consequence of viscous effects \cite{hartnoll1706} or other effects, such as Kondo physics \cite{kondo}.

\section{Hydrodynamics}
We begin by describing hydrodynamic transport in a Fermi liquid.  For simplicity, we assume an isotropic Fermi surface; new transport coefficients generically arise in the absence of rotational symmetry \cite{cook}.   In the limit where temperature $T$ is very small compared to Fermi energy $E_{\mathrm{F}}$, \emph{and} in a background magnetic field, {\color{black} we will see that } it is acceptable to neglect energy conservation and treat the only hydrodynamic degrees of freedom as charge and energy \cite{lucasreview}.   

For simplicity, we focus on flows in media which are only inhomogeneous in a single direction $y$.   We assume an isotropic fluid, so that incoherent conductivities may be neglected at low temperatures.  We expect that the resulting cartoon qualitatively captures the physics of flows in media which are inhomogeneous in both directions.   The hydrodynamic equations of charge, {\color{black} energy and }momentum conservation (up to sources) read respectively: 
\begin{subequations} \label{eq:hydro}\begin{align}
\partial_y (n v_y) &= 0, \\
\partial_y \left(T_0 s v_y - \kappa \partial_y T \right) &= 0, \\
-\partial_y (\eta \partial_y v_x + \eta_{\mathrm{H}} \partial_y v_y) &= n(E_x + Bv_y), \\
n\partial_y \mu + s\partial_y T - \partial_y ( (\zeta+\eta) \partial_y v_y - \eta_{\mathrm{H}}\partial_y v_x ) &= n(E_y - Bv_x),
\end{align}\end{subequations}
where we have approximated that $\eta$ and $\eta_{\mathrm{H}}$, the shear and Hall viscosity respectively, can be treated as constants.   Here $T_0$ is a background temperature which is independent of $y$.  For simplicity, we are neglecting the vorticity susceptibility \cite{yarom} and bulk viscosity.  We have also chosen units of charge so that the electron has charge $+1$, for convenience in what follows.
\subsection{A Narrow Channel}
We begin by briefly reviewing the scenario for negative magnetoresistance proposed in \cite{alekseev}.   Consider a long narrow channel of width $w$, inside of which is a completely homogeneous electron fluid.   We choose coordinates so that the channel is the region $\frac{1}{2}w\le |y|$.  We assume that the scattering at the boundaries is largely diffuse, in which case it is appropriate to assume no slip boundary conditions: $v_y=0$ at $y=\pm \frac{1}{2}w$.   

Suppose that we apply an electric field $E_x$, oriented down the channel.   We wish to solve (\ref{eq:hydro}) for $\mu$, $v_x$ and $v_y$, to linear order in $E_x$, in order to calculate linear response transport coefficients.  The constraint that no electric current flows through the boundary, together with charge conservation, implies that $v_y=0$.   It is straightforward to then obtain the solution to (\ref{eq:hydro}) consistent with boundary conditions: \begin{subequations}\begin{align}
v_x &= \frac{nE_x}{2\eta} \left(\frac{w^2}{4}-y^2\right), \\
\mu &= -\frac{\eta_{\mathrm{H}} E_x}{\eta}y - B \frac{nE_x}{8\eta}w^2 y + B\frac{nE_x}{6\eta}y^3.
\label{eq:muchannel}
\end{align}\end{subequations}
We have assumed for this subsection that $n$ and $\eta$ do not depend on position.    The resistivity is defined as \begin{equation}
\frac{1}{\rho} = \frac{1}{w}\int\limits_{-w/2}^{w/2} \mathrm{d}_y \frac{J_x}{E_x} = \frac{n^2 w^3}{12\eta} . \label{eq:channelres}
\end{equation}

The last fact which we need to use is that in an isotropic two-dimensional Fermi liquid, \begin{subequations}\begin{align}
\eta(B) &= \frac{\eta_0}{1+(2\omega_{\mathrm{c}}\tau_{\mathrm{ee}})^2},  \label{eq:etaB} \\
\eta_{\mathrm{H}}(B) &= 2\omega_{\mathrm{c}}\tau_{\mathrm{ee}} \eta(B), \label{eq:hallviscosity}
\end{align}\end{subequations}
where $1/r_{\mathrm{c}} = B/p_{\mathrm{F}}$ is the cyclotron radius, $\ell_{\mathrm{ee}}$ is (predominantly) the mean free path for momentum-conserving electron-electron collisions, and \begin{equation}
\eta_0 \sim np_{\mathrm{F}}\ell_{\mathrm{ee}}
\end{equation}
is an increasing function of the background density. In other words, in a two-dimensional Fermi liquid with dispersion relation $\epsilon \sim p^z$ as $p\rightarrow 0$, \begin{equation}
\ell_{\mathrm{ee}} \sim T^{-2} n^{(2z-1)/2}.  \label{eq:leen}
\end{equation}
  Combining (\ref{eq:channelres}) and (\ref{eq:etaB}), we obtain that \begin{equation}
\rho(B) = \frac{12\eta_0}{n^2w^3}\left(1- \left(\frac{\ell_{\mathrm{ee}}}{p_{\mathrm{F}}}\right)^2 B^2 + \mathrm{O}\left(B^4\right)\right).
\end{equation}
Hence we predict that there is a negative magnetoresistance.   Since $\ell_{\mathrm{ee}}\sim T^{-2}$, the effect should be more pronounced at lower temperatures (if hydrodynamics is valid).   This effect has been observed experimentally in narrow channels \cite{bandurin19}.  

A key point, however, is that (\ref{eq:muchannel}) implies the presence of a Hall voltage from $y=-w/2$ to $y=w/2$.   This means that this solution does \emph{not} immediately generalize into a continuous medium, where the fluctuating chemical potential $\mu$ must be continuous.  This means that a separate theory is required to understand transport in inhomogeneous media, which we turn to next.

\subsection{Inhomogeneous Media}
We now consider the equations (\ref{eq:hydro}) in an infinite medium with local fluid density $n(y)$.   The local viscosity $\eta$ and Hall viscosity $\eta_{\mathrm{H}}$ will also generically depend on $y$.  For simplicity, we suppose that all of these functions are periodic with some large period $L$.   Clearly, charge conservation implies that \begin{equation}
J_y = n(y)v_y(y)
\end{equation}
is a constant. 

First, let us consider applying an electric field in the $y$-direction:  $E_x = 0$ and $E_y \ne 0$.   Integrating the $x$-momentum equation over the periodic direction and denoting \begin{equation}
\langle \cdots \rangle = \frac{1}{L} \int \mathrm{d}y \cdots,
\end{equation}
we conclude that \begin{equation}
0 = \langle n(E_x + Bv_y) \rangle = BJ_y + \langle n\rangle E_x. \label{eq:hallhydro}
\end{equation} 
Hence $J_y = v_y = 0$, which implies that \begin{equation}
\partial_y (\eta \partial_y v_x) = 0.
\end{equation}
This is only satisfied for periodic functions by constant $v_x$.       The $y$-momentum equation is then only satisfied for constant $\mu$ and \begin{equation}
v_x = \frac{E_y}{B}.
\end{equation}
We conclude that \begin{subequations}\begin{align}
\sigma_{xy} &= \frac{\langle n\rangle }{B}, \\
\sigma_{yy} &= 0.
\end{align}\end{subequations}

Now we assume $E_y=0$ and $E_x\ne 0$. Then (\ref{eq:hallhydro}) implies that
 \begin{equation}
\sigma_{yx}  = \frac{J_y}{E_x} = -\frac{\langle n\rangle }{B}.
\end{equation}
For convenience in what follows, we define the periodic function $\Psi$ by the integrable equation \begin{equation}
\partial_y\Psi = n - \langle n\rangle.
\end{equation}
along with the constraint $\langle \Psi\rangle = 0$.  The $x$-momentum equation becomes \begin{equation}
-\partial_y \left(\eta \partial_y v_x + \eta_{\mathrm{H}}J_y \partial_y \frac{1}{n}\right) = E_x (n-\langle n\rangle) = E_x \partial_y \Psi,
\end{equation}
which is solved by \begin{equation}
v_x = C  - \int \mathrm{d}y \left[\frac{E_x \Psi}{\eta} + \frac{\eta_{\mathrm{H}}J_y}{\eta} \partial_y \frac{1}{n} \right].
\end{equation}
To determine the unknown constant $C$, note that \begin{align}
-BC &= \left\langle \partial_y \mu + \frac{s}{n}\partial_y T - \frac{1}{n}\partial_y ((\zeta+\eta) \partial_y v_y - \eta_{\mathrm{H}}\partial_y v_x) \right\rangle = \left\langle \frac{s}{n}\partial_y T + \partial_y \frac{1}{n} \left((\zeta+\eta) \partial_y \frac{J_y}{n} - \frac{\eta_{\mathrm{H}}}{\eta} \partial_y v_x \right)\right\rangle \notag \\
&= \left\langle \frac{s}{n}\partial_y T -E_x \frac{\langle n\rangle}{B}  \frac{\eta(\eta+\zeta)+\eta^2_{\mathrm{H}}}{\eta}\left( \partial_y \frac{1}{n}\right)^2  + \frac{\eta_{\mathrm{H}} E_x}{\eta} \Psi \partial_y \frac{1}{n} \right\rangle.
\end{align}
To fix $\partial_y T$, we use the energy conservation equation, which can be straightforwardly integrated:
\begin{equation}
\partial_y T = \frac{T_0s J_y}{n\kappa} - \frac{1}{\langle \kappa^{-1}\rangle \kappa}\left\langle  \frac{T_0s J_y}{n\kappa}  \right\rangle 
\end{equation}
(the constant is fixed by periodicity of $T$).  Lastly, to determine $\sigma_{xx}$, observe that \begin{equation}
\langle J_x\rangle = \langle (\langle n\rangle + \partial_y \Psi) v_x\rangle = C\langle n\rangle - \langle \Psi \partial_y v_x\rangle,
\end{equation}
which implies that \begin{equation}
\sigma_{xx} = \frac{\langle n\rangle^2}{B^2} \left\langle  (\eta+\zeta)\left( \partial_y \frac{1}{n}\right)^2\right\rangle + \left\langle \frac{1}{\eta} \left(\Psi - \frac{\langle n\rangle}{B} \frac{\eta_{\mathrm{H}}}{\eta}\partial_y \frac{1}{n} \right)^2 \right\rangle + \frac{\langle n\rangle ^2T_0}{B^2} \left(\left\langle \frac{s^2}{n^2\kappa} \right\rangle - \frac{1}{\langle \kappa^{-1}\rangle} \left\langle \frac{s}{n\kappa}\right\rangle^2 \right).
\end{equation}

Clearly, we have found a positive semidefinite conductivity tensor as required on physical grounds.  It is straightforward to convert to a resistivity matrix: \begin{subequations}\begin{align}
\rho_{xx} &= 0, \\
\rho_{xy} &= -\rho_{yx} = -\frac{B}{\langle n\rangle}, \\
\rho_{yy}&= \left\langle (\eta+\zeta) \left(\partial_y \frac{1}{n}\right)^2 + \frac{1}{\eta} \left(\frac{B}{\langle n\rangle}\Psi - \frac{\eta_{\mathrm{H}}}{\eta}\partial_y \frac{1}{n} \right)^2 \right\rangle + T_0 \left(\left\langle \frac{s^2}{n^2\kappa} \right\rangle - \frac{1}{\langle \kappa^{-1}\rangle} \left\langle \frac{s}{n\kappa}\right\rangle^2 \right).
\end{align}\end{subequations}

\begin{figure}[t]
\centering
\includegraphics[width=3in]{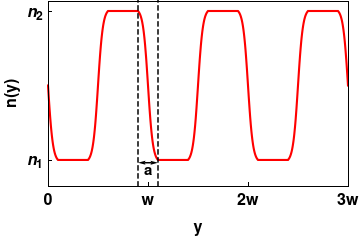}
\caption{A sketch of $n(y)$ in a periodic system.  $n(y)$ is approximately $n_1$ in half of the channel (length $w/2$) and approximately $n_2$ in the other half.   The transition region between the two ``domains" is of length $a$.  We assume $a\ll w$ and $n_1 \ll n_2$. }
\label{fig:cartoon}
\end{figure}

It remains to check whether the magnetoresistance can be negative.  {\color{black} We first explain that, consistent with \cite{levchenko1, levchenko2}, the thermal conductivity can be neglected.   (\emph{i}) At low temperature in a Fermi liquid, $\kappa \sim T_0^{-1}$ and $s_0 \sim T$, while $\eta \sim T^{-2}$.  Hence, the correction to the hydrodynamic resistivity arising from the thermal conductivity is suppressed by a power of $T^6$, which is quite small at low temperatures.  (\emph{ii}) At very long wavelengths, but finite temperature, the dominant term in $\rho_{yy}$ is the $\Psi$-dependent term, which diverges with the wavelength of the inhomogeneity in the charge density; in contrast, the contribution to $\rho_{yy}$ coming from the thermal conductivity does not diverge with the length scale of the inhomogeneity.}  

For the rest of this section, we focus our study on the cartoon system shown in Figure \ref{fig:cartoon}.   From this cartoon, and (\ref{eq:hallviscosity}), we can estimate that \begin{subequations}\begin{align}
\left\langle \frac{1}{\eta} \left(\frac{B}{\langle n\rangle}\Psi - \frac{\eta_{\mathrm{H}}}{\eta}\partial_y \frac{1}{n} \right)^2 \right\rangle &\sim  \frac{1}{\eta(n_1,B)} \frac{B^2}{n_2^2} (n_2w)^2 \sim \frac{(Bw)^2}{\eta(n_1,B)}, \\
 \left\langle \eta \left(\partial_y \frac{1}{n}\right)^2 \right\rangle \sim \frac{a}{w} \frac{\eta(n_1,B) }{n_1^4} \left(\frac{n_2}{a}\right)^2 .
\end{align}\end{subequations}
As $B\rightarrow 0$, we can estimate that \begin{equation}
\frac{\partial \rho_{yy}}{\partial B^2} \sim - \frac{n_2^2 \ell_{\mathrm{ee,1}}^2}{n_1^4aw p_{\mathrm{F},1}^2} \eta(n_1,0)  + \frac{w^2}{\eta(n_1,0)}.
\end{equation}
Hence, magnetoresistance is negative when \begin{equation}
aw^3 \ll \left(\frac{n_2}{n_1} \frac{\eta(n_1) \ell_{\mathrm{ee,1}}}{n_1 p_{\mathrm{F,1}}}\right)^2 \sim \left(\frac{n_2}{n_1}  \ell_{\mathrm{ee,1}}^2\right)^2 \label{eq:NMRcondition}
\end{equation}

Of course, hydrodynamics itself is only valid when $w\gg a \gg \ell_{\mathrm{ee}}$, where the last inequality should (conservatively) hold for the maximal value which $\ell_{\mathrm{ee}}$ takes in the domain.   Assuming (\ref{eq:leen}), we find that (\ref{eq:NMRcondition}) becomes $n_2^{4z-4} \ll n_1^{4z-4}$, which implies $z<1$.   We do not know of any physical systems with $z<1$, so this argument suggests that even the cartoon model above, which is perhaps absurd for a realistic metal, is insufficient to lead to negative magnetoresistance at small $B$.

At large $B$, using (\ref{eq:etaB}), we instead have: 
\begin{equation}
\rho_{yy} \sim \frac{B^4w^2}{\eta(n_1,0)} \frac{\ell_{\mathrm{ee,1}}^2}{p_{\mathrm{F,1}}^2} + \frac{n_2^2}{awn_1^4} \eta(n_1,0)\frac{p_{\mathrm{F,1}}^2} {\ell_{\mathrm{ee,1}}^2B^2},
\end{equation}
which exhibits positive magnetoresistance whenever \begin{equation}
B\gtrsim \frac{p_{\mathrm{F,1}}}{\ell_{\mathrm{ee,1}}^{1/3} \ell_{\mathrm{ee,2}}^{2/3}},  \label{eq:Bineq}
\end{equation}
assuming $a,w \gg \ell_{\mathrm{ee,2}}$.   As (\ref{eq:Bineq}) is a sufficiently small magnetic field to estimate the magnetoresistance by the $B^2$-correction to resistivity, we conclude that for any value of $B$, magnetoresistance will generally be positive, even in highly inhomogeneous metals.

The one shortcoming in our argument is, of course, that the system was only inhomogeneous in one of the two directions.  However, we do not expect that a fully two-dimensional calculation would qualitatively change the physics described above.  In the presence of a magnetic field, it is not possible to push the electron fluid along contours of almost zero resistance due to local Hall effects.   We leave a final resolution of the two-dimensional transport problem to elsewhere.

{\color{black} 
To summarize our findings thus far, we have seen that even in highly inhomogeneous metals with sharp ``domain walls" between regimes of different density, the magnetoresistance remains positive.  This is despite our attempt to engineer a flow through a narrow channel analogous to \cite{alekseev}.  The difference between our calculation and that of \cite{alekseev} is that, as mentioned previously, in the narrow channel of \cite{alekseev} there is a Hall voltage between the two sides of the channel while no current flows between them ($J_y=0$), while in a continuous medium the voltage must be continuous and a current flows ($J_y\ne 0$).   This qualitative change to the flow of current creates additional magnetic field dependent corrections to transport which cause a large positive magnetoresistance \cite{levchenko1, levchenko2} to the electrical resistivity governing bulk transport.  It is, however, plausible that a negative magnetoresistance could be seen in narrow channels, such as in a recent experiment in graphene \cite{bandurin19}.

}

 \section{Kinetic Theory}
\label{sec:kinetic} 
 
 Next, we ask whether it is possible to have negative magnetoresistance at the \emph{onset} of hydrodynamic behavior at ultra low temperatures, below which the physics is described by essentially free quasiparticles moving through a random medium.
 
 \subsection{Weakly Inhomogeneous Media}
 To begin, we briefly recall some known results from the theory of transport in weakly inhomogeneous metals \cite{lucasMM}.   Such results can be derived from kinetic theory as well \cite{hartnoll1706, lucas1711}, though we will not do so here.   Consider an arbitrary quantum many-body system (not only a Fermi liquid with well-defined quasiparticles) with a conserved U(1) charge, whose (effective) Hamiltonian $H_0$ is translation invariant \emph{in the continuum} and hence momentum conserving.   Suppose that the low energy theory is described by Hamiltonian \begin{equation}
 H = H_0 - \int\mathrm{d}^d\mathbf{x} \; \mu(\mathbf{x}) n(\mathbf{x}),
 \end{equation}
 where $n(\mathbf{x})$ is the charge density operator and $\mu(\mathbf{x})$ is a perturbatively small inhomogeneous coefficient.   Then the electrical resistivity tensor $\rho_{ij}$ is \begin{equation}
\rho_{ij} = \frac{1}{n_0^2}\int \frac{\mathrm{d}^d\mathbf{k}}{(2\mpi)^d} k_i k_j |\mu(\mathbf{k})|^2 \times \mathcal{A}_{nn}(\mathbf{k})    + \mathrm{O}(\mu^3) ,
 \end{equation}
 where $n_0 = \langle n(\mathbf{x})\rangle_{H_0}$ is the average charge density, $\mu(\mathbf{k})$ is the Fourier transform of $\mu(\mathbf{x})$, up to an overall coefficient related to the volume of spacetime, and
 \begin{equation}
 \mathcal{A}_{nn}(\mathbf{k}) = \lim_{\omega \rightarrow 0} \frac{\mathrm{Im}\left(G^{\mathrm{R}}_{nn}(\mathbf{k},\omega)\right)}{\omega}. \label{eq:Annk}
 \end{equation}
 is the spectral weight of the charge density operator at wave number $\mathbf{k}$.
 Details of this derivation can be found in \cite{lucasMM}.  {\color{black} This result can be understood as a generalization of the Born approximation for electron-impurity scattering to generic interacting systems scattering off of arbitrary kinds of weak disorder.}

\subsection{Kinetic Model}
All we need to do in order to calculate resistivity is to evaluate $\mathcal{A}_{nn}(\mathbf{k})$, and towards this end we use kinetic theory as a toy model for the spectral weight of the density operator across the ballistic-to-hydrodynamic crossover.   We restrict our focus to a toy model of the kinetic theory of a two-dimensional Fermi liquid \cite{molenkamp, levitov1607, hartnoll1706, lucas1612, levitov1612}, whose properties have been extensively studied in these previous papers.  Here we simply review what the model is and how to solve it.  Let \begin{equation}
f(\mathbf{x},\mathbf{p}) = f_{\mathrm{eq}}(\mathbf{p}) + \mdelta f(\mathbf{x},\mathbf{p})
\end{equation} be the distribution function of the fermionic quasiparticles, with $f_{\mathrm{eq}}$ the Fermi-Dirac distribution and $\mdelta f$ a perturbatively small correction due to the presence of the external electric field.     The linearized Boltzmann equation reads \begin{equation}
\partial_t \mdelta f + \mathbf{v}\cdot \partial_{\mathbf{x}} \mdelta f +\mathbf{E}\cdot \partial_{\mathbf{p}} f_{\mathrm{eq}} + (\mathbf{v}\times\mathbf{B})\cdot \partial_{\mathbf{p}} \mdelta f = -\mathsf{W}\otimes \mdelta f,
\end{equation}where $\mathsf{W}$ corresponds (abstractly) to the linearized collision operator.  {\color{black} Note that in a background magnetic field, the global momentum is not a conserved quantity due to the Lorentz force; the fact that electron-electron collisions conserve momentum alone implies that the linearized collision operator $\mathsf{W}$ does not relax momentum. }   As explained in \cite{lucasreview}, at very low temperatures in a Fermi liquid, we may approximate that $f_{\mathrm{eq}} = \mathrm{\Theta}(\epsilon_{\mathrm{F}} - \epsilon(\mathbf{p}))$ and that \begin{equation}
\mdelta f = \mdelta(\epsilon_{\mathrm{F}} - \epsilon(\mathbf{p})) \times \Phi(\mathbf{x},\theta),
\end{equation}
where the variable $\Phi$ is defined only on the Fermi surface.  In a rotationally invariant model, the Fermi surface is easily parameterized by  an angle $\theta$.

For simplicity, we will {\color{black} begin by} assuming a relaxation time approximation for the linearized collision integral.   Defining \begin{subequations}\begin{align}
\Phi &= \sum_{n\in\mathbb{Z}} a_n \mathrm{e}^{\mathrm{i}n\theta}, \\
\mathbb{P}\Phi &= \sum_{|n|\ge 2} a_n \mathrm{e}^{\mathrm{i}n\theta},
\end{align}\end{subequations}
we approximate that \begin{equation}
\partial_t \Phi + \mathbf{v}\cdot \partial_{\mathbf{x}} \Phi + \mathbf{E}\cdot \mathbf{v} + \omega_{\mathrm{c}} \partial_\theta \Phi = -\frac{1}{\tau_{\mathrm{ee}}}\mathbb{P}\Phi , \label{eq:relboltz}
\end{equation}
where the cyclotron frequency is
\begin{equation}
\omega_{\mathrm{c}} = \frac{v_{\mathrm{F}}B}{p_{\mathrm{F}}}.
\end{equation}
Despite the fact that the collision integral is not quantitatively accurate \cite{ledwith1, ledwith2, levitov19}, this model is exactly solvable for many purposes, including ours.  As discussed in the following section, we expect that the qualitative physics described below remains relevant for realistic metals close to the hydrodynamic regime.

It is useful for us to recast the cartoon Boltzmann equation above as follows.  Denote $\Phi(\theta)$ as \begin{equation}
|\Phi\rangle = \sum_{n\in\mathbb{Z}} a_n |n\rangle,
\end{equation}
where we define the inner product \begin{equation}
\langle n|m\rangle = \nu\,, \mdelta_{nm},
\end{equation}
with $\nu$ the density of states of the Fermi liquid. Defining the matrices \begin{subequations}\begin{align}
\mathsf{L}(\nabla) |n\rangle &= \frac{v_{\mathrm{F}}}{2} (\partial_x + \mathrm{i}\partial_y) |n-1\rangle + \frac{v_{\mathrm{F}}}{2} (\partial_x - \mathrm{i}\partial_y) |n+1\rangle + \mathrm{i}n\omega_{\mathrm{c}}|n\rangle , \\
\mathsf{W}|n\rangle &= \frac{1}{\tau_{\mathrm{ee}}} (1-\mdelta_{n,0} - \mdelta_{n,1} - \mdelta_{n,-1}) |n\rangle,
\end{align}\end{subequations}
the spectral weight $\mathcal{A}_{nn}(\mathbf{k})$ is computed in kinetic theory as \cite{hartnoll1706}:
\begin{equation}
\mathcal{A}_{nn}(\mathbf{k}) = \langle 0| ( \mathsf{W} + \mathsf{L}(\mathbf{k}))^{-1}|0\rangle . \label{eq:kineticA}
\end{equation}
As our model is rotationally invariant, we can set $\mathbf{k} = k\hat{\mathbf{x}}$ when evaluating (\ref{eq:kineticA}), without loss of generality.

\subsection{Spectral Weight}
In order to evaluate (\ref{eq:kineticA}), we use a simple trick from \cite{levitov1607}.   Let \begin{equation}
\mathsf{G}(\mathbf{k}) = (\mathsf{W} + \mathsf{L}(\mathbf{k}))^{-1}.
\end{equation}  Suppose we can exactly evaluate \begin{equation}
\mathsf{G}_0(\mathbf{k}) = \left(\frac{1}{\tau_{\mathrm{ee}}} + \mathsf{L}(\mathbf{k})\right)^{-1},
\end{equation}
where the first term in the inverted matrix above implicitly multiplies the identity matrix.   Denoting $\mathsf{X}$ as the projection onto the three $|n|\le 1$ harmonics, and observing that \begin{equation}
\mathsf{G} = \left(\mathsf{G}_0^{-1} - \frac{1}{\tau_{\mathrm{ee}}}\mathsf{X}\right)^{-1},  \label{eq:GG0}  
\end{equation}
we obtain a simple formula:  if $\tilde{\mathsf{G}}$ denotes the $3\times 3$ submatrix of $\mathsf{G}$ corresponding to the $|n|\le 1$ harmonics, and similarly for $\tilde{\mathsf{G}}_0$: \begin{equation}
\tilde{\mathsf{G}} = \left(1 - \frac{1}{\tau_{\mathrm{ee}}} \tilde{\mathsf{G}}_0 \right)^{-1} \tilde{\mathsf{G}}_0. \label{eq:trick}
\end{equation}
{\color{black} These results follow from block matrix inversion identities.  We only need the $3\times 3$ subspace of harmonics because our goal is to calculate the spectral weight $\mathcal{A}_{nn}$, which is given by (\ref{eq:kineticA}) and thus  $\langle 0| \tilde{\mathsf{G}}|0\rangle$.}

Now, we evaluate $\tilde{\mathsf{G}}_0$.   To do so, we find the eigenvectors of $\mathsf{L}$, and it is easiest to do so by temporarily reverting back to the $\theta$-basis.   With $\mathbf{k}=k\hat{\mathbf{x}}$, we find \begin{equation}
\left(\mathrm{i}kv_{\mathrm{F}}\cos\theta - \omega_{\mathrm{c}}\partial_\theta \right)\Phi_\lambda(\theta) =\lambda\Phi_\lambda(\theta)
\end{equation}
with $\lambda$ an eigenvalue to be determined.   A simple calculation \cite{hedegard} shows that \begin{equation}
\Phi_\lambda(\theta) = \exp\left[\mathrm{i}\tilde n\theta + \mathrm{i}\frac{kv_{\mathrm{F}}}{\omega_{\mathrm{c}}} \sin \theta \right], \;\;\; \lambda = -\mathrm{i}\tilde n\omega_{\mathrm{c}}.
\end{equation}
In what follows, we define $\ell_B = v_{\mathrm{F}}/\omega_{\mathrm{c}}$ and $\ell_{\mathrm{ee}} = v_{\mathrm{F}}\tau_{\mathrm{ee}}$ for simplicity.   Since \begin{equation}
\tilde{\mathsf{G}}_0 = \sum_{\tilde n \in \mathbb{Z}} \frac{|\Phi_{\tilde n}\rangle\langle \Phi_{\tilde n}|}{\tau_{\mathrm{ee}}^{-1}-\mathrm{i}\tilde n\omega_{\mathrm{c}}},
\end{equation}
and \begin{equation}
\frac{\langle \Phi_{\tilde n}|n\rangle}{\nu} = \int\limits_0^{2\mpi}\frac{\mathrm{d}\theta}{2\mpi} \mathrm{e}^{\mathrm{i}(n-\tilde n ) \theta - \mathrm{i}k\ell_B\sin\theta}  = \mathrm{J}_{n-\tilde n}(k\ell_B),
\end{equation}
we conclude that \begin{equation}
\frac{\langle n^\prime|\tilde{\mathsf{G}_0}|n\rangle}{\nu} = v_{\mathrm{F}} \sum_{\tilde n \in \mathbb{Z}} \frac{\mathrm{J}_{n^\prime - \tilde n}(k\ell_B)\mathrm{J}_{n - \tilde n}(k\ell_B) }{\ell_{\mathrm{ee}} - \mathrm{i}\tilde n \ell_B}. \label{eq:Bres}
\end{equation}

{\color{black} Combining (\ref{eq:kineticA}) (\ref{eq:trick}) and (\ref{eq:Bres}) we find an explicit formula for $\mathcal{A}_{nn}$.  Then we may use (\ref{eq:Annk}), along with a physically sensible choice for the inhomogeneity $\mu(\mathbf{k})$, to numerically evaluate the resistivity.   Note that the inhomogeneity $\mu(\mathbf{k})$ is not a property of the kinetic theory computation, but is a property of the environment in which the electrons are moving.}

    A physically sensible choice is to assume random point-like charged impurities, placed at a distance $\xi$ ``above the plane" in which the electrons flow.   (These are analogous to impurities in the gates of a heterostructure).   One finds that \cite{hartnoll1706}
\begin{align}
\left | \mu (k) \right |^2 \propto \frac{\mathrm{e}^{-2\,k \,\xi}}
{\left(  k+k_{\mathrm{TF}}\right)^2 } \,,
\label{eqimp}
\end{align}
where $ k _{\mathrm{TF}}$ is a Thomas-Fermi screening wave number.

	\begin{figure}[t]
\includegraphics[width=0.45\linewidth]{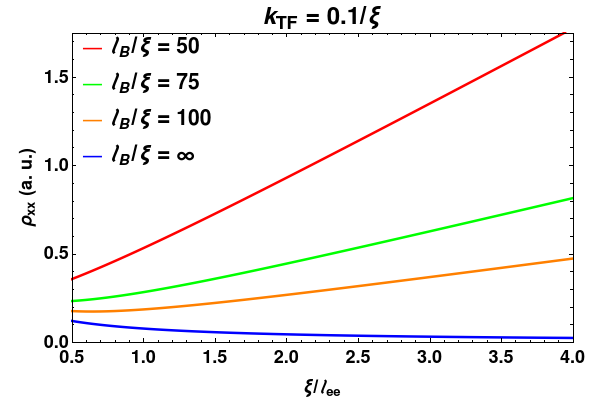} \quad
\includegraphics[width=0.45\linewidth]{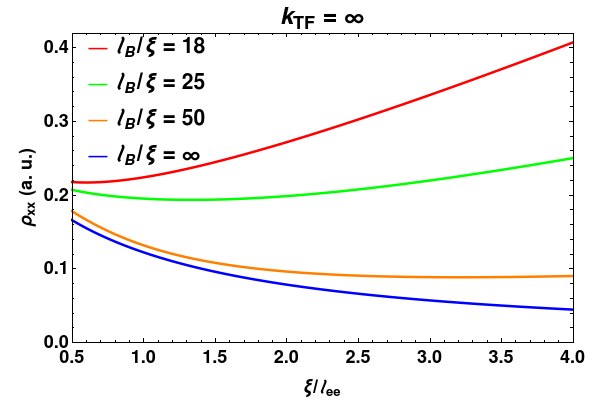}
\caption{\label{fig1}Plots of resistivity as a function of $\xi/\ell_{\mathrm{ee}} $, for fixed values of $\ell_B/\xi$.}
\end{figure}

In Figure \ref{fig1}, we show the resistivity $\rho$ as a function of $\xi/\ell_{\mathrm{ee}}$.  Using (\ref{eq:leen}), we can interpret this as $\rho$ as a function of $T^2$ (up to logarithms) in a Fermi liquid.  Observe that when $B=0$ ($\ell_B=\infty$) that $\rho$ is a \emph{decreasing function of temperature}.  This is the manifestation of the Gurzhi effect seen in a bulk crystal \cite{hartnoll1706}, and when we consider the limit $\ell_B\rightarrow \infty$, we exactly recover the results of \cite{hartnoll1706}.    Rather surprisingly, we find that a cyclotron radius $\ell_B> 10 \xi$ essentially leads to $\partial \rho/\partial T>0$ at all temperatures.   In other words, in a bulk crystal (unlike in a narrow channel), the Gurzhi effect is exquisitely sensitive to a magnetic field.   Hence, a simple experimental test for viscous origins of a resistance minimum is to simply study the sensitivity of the effect to a small external magnetic field.

\begin{figure}[t]
\includegraphics[width=0.45\linewidth]{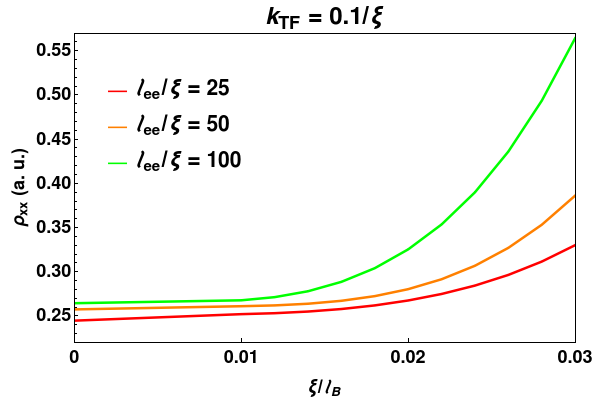} \quad
\includegraphics[width=0.45\linewidth]{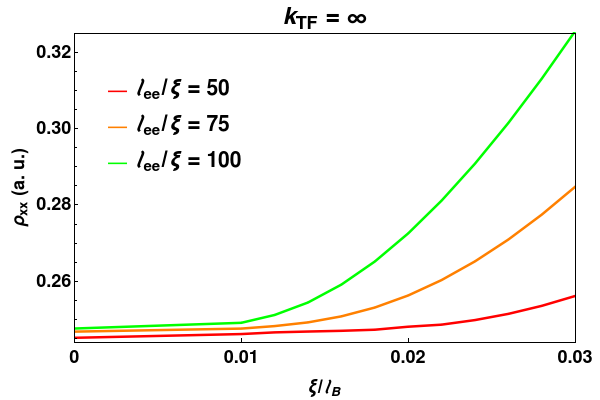}
\caption{\label{fig2}Plots of resistivity as a function of $\xi/\ell_B $, for fixed values of $\ell_{\mathrm{ee}} /\xi$.}
\end{figure}

In Figure \ref{fig2}, we show $\rho$ as a function of $B$ at various temperatures.   It is clear that regardless of the choice of parameters, $\rho(B)$ is a rapidly increasing function.

While we have not demonstrated that magnetoresistance is positive across the ballistic-to-hydrodynamic crossover for large amplitude inhomogeneity, we see no reason for this to not be the case.

\subsection{A More Sophisticated Model}
{\color{black}
Lastly, let us revisit the simplified model of kinetic theory discussed in the previous section.   It has been pointed out in \cite{levitov19} that when the temperature $T\ll T_{\mathrm{F}}$ (the Fermi temperature) in a low temperature Fermi liquid with a circular Fermi surface, analogous to our model above, that $\langle n|\mathsf{W}|n\rangle \sim (T/T_{\mathrm{F}})^4 n^4 $ when $n$ is odd, while $ \langle n|\mathsf{W}|n\rangle \sim (T/T_{\mathrm{F}})^2$ when $n$ is even.  (We have ignored logarithmic factors in both $T/T_{\mathrm{F}}$ and $n$, for convenience.)    

One simple model which retains some of the simplicity of the relaxation time approximation, but at the same time includes the hierarchy of scattering times described above, is as follows:  choose $n_{\mathrm{max}} \ge 3$ to be an odd harmonic.   Then define the collision integral $\mathsf{W}$ to be \begin{equation}
\langle n|\mathsf{W}|n^\prime \rangle = \frac{\mdelta_{nn^\prime}}{\tau_{\mathrm{ee}}} \times \left\lbrace \begin{array}{ll} 0 &\  |n| \le 1 \\ 1 &\ n=\pm 2, \pm 4, \ldots \\ 1 &\
\pm (n_{\mathrm{max}}+2),  \pm (n_{\mathrm{max}}+4), \ldots \\ n^4/(n_{\mathrm{max}}+2)^4  &\ n = \pm 3, \ldots, \pm n_{\mathrm{max}}.    
\end{array}\right..  \label{eq:complicatedW1}
\end{equation}
We say that the model with $n_{\mathrm{max}}=1$ is the relaxation time model from above, with $\mathsf{W}$ given by (\ref{eq:relboltz}).   The parameter $n_{\mathrm{max}}$ can be thought of as a simple proxy for $T/T_{\mathrm{F}}$, with \begin{equation}
n_{\mathrm{max}} \sim \sqrt{\frac{T_{\mathrm{F}}}{T}}.
\end{equation}
Metals such as graphene exhibit signatures of hydrodynamic flow when $T/T_{\mathrm{F}}\sim 0.1$, so we expect that $n_{\mathrm{max}}=3$ or $n_{\mathrm{max}}=5$ are reasonable models for this system, as well as other low density Fermi liquids which might be near the hydrodynamic regime in experimental devices.

An even simpler model, which we present for illustrative purposes and to demonstrate that our results are not finely tuned, is to simply assume that all odd harmonics with $|n| \le n_{\mathrm{max}}$ are conserved quantities: \begin{equation}
\langle n|\mathsf{W}|n^\prime \rangle = \frac{\mdelta_{nn^\prime}}{\tau_{\mathrm{ee}}} \times \left\lbrace \begin{array}{ll} 0 &\  |n| \le 1 \\ 1 &\ n=\pm 2, \pm 4, \ldots \\ 1 &\
\pm (n_{\mathrm{max}}+2),  \pm (n_{\mathrm{max}}+4), \ldots \\ 0  &\ n = \pm 3, \ldots, \pm n_{\mathrm{max}}.
 \end{array}\right.. \label{eq:complicatedW2}
\end{equation}

Regardless of the collision integral that we choose, we solve this problem the same way as before.    We evaluate (\ref{eq:GG0}) with \begin{equation}
\mathsf{X} = 1 - \tau_{\mathrm{ee}}\mathsf{W}
\end{equation}
denoting the components of the collision integral which deviate from the identity (up to a factor of $\tau_{\mathrm{ee}}$.   As in our earlier model, $\mathsf{X}$ is only non-vanishing in a finite number of rows and columns.   Hence, when we evaluate $\mathcal{A}_{nn}$ using (\ref{eq:kineticA}), we need only evaluate a finite dimensional block of (\ref{eq:GG0}).  This makes the problem numerically tractable.

%


	\begin{figure}[t]
\includegraphics[width=0.45\linewidth]{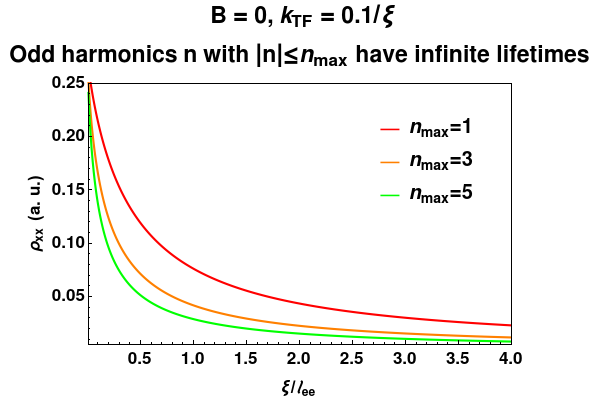} \quad
\includegraphics[width=0.45\linewidth]{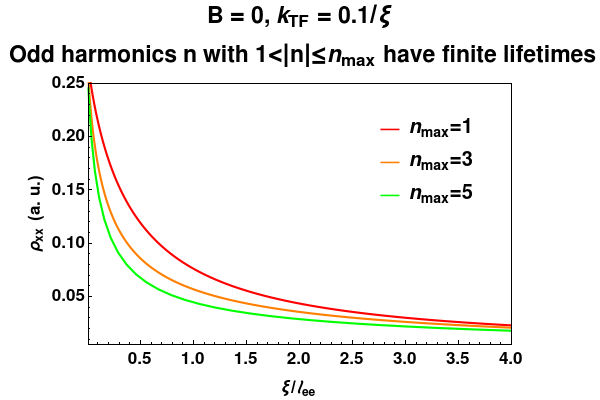}
\caption{\label{fig1}Plots of resistivity as a function of $\xi/\ell_{\mathrm{ee}} $ for $B=0 $, for models with odd harmonics up to $ |n| \leq n_{\text{max}}$ included.   Left: the model (\ref{eq:complicatedW2}); right: the model (\ref{eq:complicatedW1}).}
\label{fig:B0complicated}
\end{figure}

\begin{figure}[t]
\includegraphics[width=0.45\linewidth]{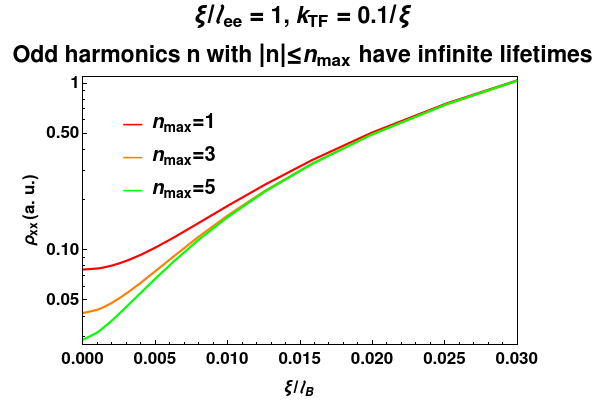}\quad
\includegraphics[width=0.45\linewidth]{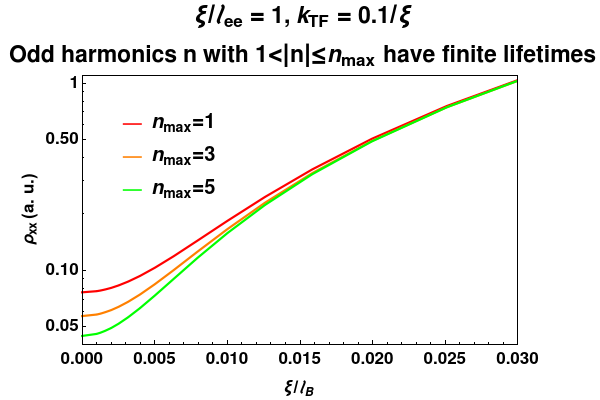}
\caption{\label{fig4}Plots of resistivity as a function of $\xi/\ell_B $, for fixed values of $\ell_{\mathrm{ee}} /\xi$, for models with odd harmonics $ |n| \leq n_{\text{max}} $ included. Left: the model (\ref{eq:complicatedW2}); right: the model (\ref{eq:complicatedW1}).}
\label{fig:Bcomplicated}
\end{figure}

In Figure \ref{fig:B0complicated}, we show the resistivity as a function of $\xi/\ell_{\mathrm{ee}}$ (a proxy for $T^2$, as before).  While the curves are qualitatively similar regardless of $n_{\mathrm{max}}$ or the precise collision integral used, there are notable quantitative differences between the curves.  Figure \ref{fig:Bcomplicated} shows that after turning on even a very small magnetic field, the quantitative discrepancies between these different models largely disappear.   Just as importantly, the trend of positive magnetoresistance persists and is not sensitive to the details of the microscopic collision operator.   Hence, our conclusion that magnetoresistance in bulk crystals is positive at the onset of the hydrodynamic regime is not an artifact of the relaxation time approximation; rather, it is a generic feature of interacting two-dimensional Fermi liquids.
}

\section{Conclusion}
\label{conclude}

In this paper, we have extended the arguments of \cite{levchenko1, levchenko2} and demonstrated that magnetoresistance is essentially always positive in simple electron liquids with strong interactions, regardless of the strength of inhomogeneity, and even when interactions are not very strong.   We expect that our most relevant observation is the extreme sensitivity of the resistivity to an external magnetic field in or near a hydrodynamic flow regime of the electrons.   We expect that this large positive magnetoresistance can serve as an experimental test for the Gurzhi effect, when manifested as a resistance minimum, in bulk transport measurements.

\section*{Acknowledgements}
AL was supported by the Gordon and Betty Moore Foundation via grant GBMF4302.
 


\bibliographystyle{unsrt}
\addcontentsline{toc}{section}{References}
\bibliography{viscousbib}

\end{document}